\documentclass[prd,nofootinbib,twocolumn,showpacs]{revtex4}  
\usepackage{mathrsfs}
\usepackage{graphicx}
\usepackage{bm}
                                                                                                                                      
\begin{document}
\title{A note  on the stability of axionic D-term s-strings}
                                                                                                                                      
\author{Ana Ach\'ucarro} \affiliation{Lorentz Institute of Theoretical
Physics, University of Leiden, The Netherlands}
\affiliation{Department of Theoretical Physics, The University of
the Basque Country UPV-EHU, Bilbao, Spain}

\author{Kepa Sousa}
\affiliation{Lorentz Institute of Theoretical Physics, University of Leiden, The Netherlands}
                                                                                                                                      
\begin{abstract}
We investigate the stability of a new class of BPS cosmic strings in
N=1 supergravity with D-terms recently proposed by Blanco-Pillado,
Dvali and Redi.  These have been conjectured to be the low energy
manifestation of D-strings that might form from tachyon condensation
after D- anti-D-brane annihilation in type IIB superstring
theory. There are three one-parameter families of cylindrically
symmetric one-vortex solutions to the BPS equations (tachyonic,
axionic and hybrid). We find evidence that  
the zero mode in the axionic case, or $s-$strings, can be excited. Its evolution leads to the decompactification of four-dimensional spacetime at late times, with a rate that decreases with decreasing brane tension.
 \end{abstract}

\pacs{11.27.+d, 11.25.Mj, 04.65.+e, 12.60.Jv}

\maketitle

Recently Blanco-Pillado, Dvali and Redi have proposed a model to
describe a D-brane anti-D-brane unstable system after compactification to four
dimensions \cite{BDR}. In the type IIB case, the tension of the branes
appears in the four-dimensional effective theory as a constant Fayet
Iliopoulos term which allows for the existence of non-singular BPS
axion-dilaton strings generalising earlier work by \cite{HN,BDP,DBD} in
heterotic scenarios. \\

It was shown that this model contains three different families of BPS
cosmic string solutions: {\it tachyonic} or $\phi$-strings, with
winding $n$ (see also \cite{DKP}), {\it axionic}\footnote{The name
axionic is somewhat misleading for the solutions discussed here since
there is no $a F {\tilde F}$ coupling, in particular they do not share the features usually associated with axionic strings \cite{HN}.} or $s$-strings with winding $m$
and a third type which is the special case where the following
relation is satisfied: $q |m| = |n|$. We are using the same notation
of \cite{BDR}.  We call the third type {\it hybrid} strings. Each of
the families is parametrized by a real number, $\kappa$, defined by
the relation:
\begin{equation}
s = 2 \frac{\delta}{q} (|n|-q|m|)\log r - 2 \frac{\delta}{q}\log |\phi| + \kappa. 
\label{zeromode}
\end{equation}
In the first two families the parameter $\kappa$ is associated to a zero mode
that connects vortex solutions with different core radius and equal
magnetic flux, as can be seen by setting:
\begin{equation}
k=2 \frac{\delta}{q} (|n|-q|m|)\log R
\end{equation}
In this way $R$ gives the scale of the core radius.  
In the third case the same parameter measures which of
the fields, the dilaton or the tachyon, contributes more to compensate
the Fayet-Iliopoulos term.\\

Supergravity effects were considered in \cite{BDR} and, as expected on general grounds
\cite{UAD,DHKLZ}, the zero mode survives the coupling to supergravity.\\

The $s$-strings are peculiar . The authors of \cite{BDR} argued that
they should be associated with D anti-D bound states that are unstable
in ten dimensions, and therefore only exist after compactification.
They also noted that they share some features with semilocal strings
in the Bogomolnyi limit \cite{AV}.  In the semilocal case any
excitation of the zero mode \cite{H} leads invariably to the spread of
the magnetic field and the eventual disappearance of the strings
\cite{leese}.\\

We have studied numerically the dynamics of this zero mode and we find
that also in the $s$-string case it can be excited and will lead to
the dissolution of the strings. Since in this process the field $s$
grows without bound, and therefore also the compactification volume
modulus, our result would appear to imply the decompactification of
spacetime from four to ten dimensions at late times.\\

We start by briefly reviewing the model of \cite{BDR} and the three
families of BPS strings. We then analyse the zero mode numerically and
conclude that it can be excited. For comparison, we show the results
of the same analysis on the $\phi$-string, where a perturbation of the
string leads to some oscillations but no runaway behaviour.

\section{The model}

The authors of \cite{BDR} proposed a supersymmetric abelian Higgs
model containing a vector superfield $V$ and $N$ chiral
superfields. Here we only need to consider one chiral superfield
$\Phi$ with charge $q$, that represents the tachyon.  The model also
involves an axion-dilaton superfield $S$ which is coupled to the gauge
multiplet in the usual way. Its lowest component is $s+ia$ where the
axion $a$ is the four-dimensional dual of the Ramond-Ramond two-form
zero mode after compactification. Its scalar partner $s$ is some
combination of the dilaton and the volume modulus.  We take the  K\"ahler form 
to be $ K = -M_p^2 \log ( S + \bar S)$ and the gauge kinetic
function are set to be constant, $f(S)=1/g^2$, where $g$ is the gauge coupling.
With this choice, there is no $a F_{\mu\nu} \tilde F^{\mu\nu} $ term
in the action. The $U(1)$ symmetry is not anomalous, it is the
diagonal combination of the $U(1)$s from the D- anti D system.

In component notation the bosonic sector of the lagrangian, after eliminating the auxiliary
field from the vector multiplet, is:
\begin{eqnarray}
\mathscr{L}&=& -|D_\mu \phi|^2 - K_{S \bar S} |D_\mu S|^2 - \scriptstyle 
\frac{1}{4} \displaystyle g^{-2} F^{\mu \nu } F_{\mu \nu}{}
\nonumber \\
&& {} -\scriptstyle \frac{1}{2}  \displaystyle  g^2 ( \xi + 2 \delta K_S - q |\phi|^2)^2   .
\end{eqnarray}
$K_S$ and $K_{ S \bar S}$ represent the derivatives of the Kahler potential respect
to the fields $S$ and $\bar S$. 
$\phi$ is the lowest component of the chiral
field field, $A_\mu$ is the $U(1)$ gauge field, and $F_{\mu \nu} =
\partial_\mu A_\nu - \partial_\nu A_\mu$ the associated abelian field
strength,
 \begin{equation}
D_\mu \phi = \partial_\mu \phi - i q A_\mu \phi, \qquad D_\mu S = \partial_\mu S + i 2 \delta A_\mu . 
\end{equation}
$\delta$ is the coupling of the axion to the gauge field. In the case
of anomalous $U(1)$, where f(S) = S, it is called the Green-Schwarz
parameter and is fixed by the value of the anomaly. Here $\delta$ is
not determined. \\
 
It is convenient to write the bosonic part of the lagrangian using the
rescalings: {\setlength\arraycolsep{4pt}
\begin{eqnarray}
\phi = \sqrt{\xi/q} \; \hat \phi & s= \delta M_p^2/\xi \; \hat s   & a = 2 \delta/q \; \hat a 
\nonumber \\
A_\mu = g \sqrt{\xi/q} \; \hat  A_\mu & x=(g\sqrt{\xi q} )^{-1} \; \hat x  \ ,
\end{eqnarray}}
With these definitions the axion $\hat a$ is defined modulo $2 \pi$,
and $\delta$ is rescaled away. After dropping the hats, the 
bosonic sector of the lagrangian reads:
{\setlength\arraycolsep{2pt}
\begin{eqnarray}
\mathscr{L}(\xi g)^{-2}  &=& -|D_\mu \phi|^2-\scriptstyle \frac{1}{4} \displaystyle ( \alpha s)^{-2}
(\partial_\mu s)^2  - (\alpha /s)^{2}(\partial_\mu a +A_\mu)^2  {}
\nonumber \\
&& {}- \scriptstyle \frac{1}{4} \displaystyle F^{\mu \nu} F_{\mu \nu} -\scriptstyle  \frac{1}{2} 
\displaystyle (1- s^{-1} - |\phi|^2)^2  \ ,
\label{bosonic_lag2}
\end{eqnarray}} 
with $D_\mu \phi = \partial_\mu \phi - i A_\mu \phi $, and
$\alpha^2=\xi/(q M_p^2)$.
Note that, since $\alpha$ is the symmetry breaking scale in Plank
units, ignoring supergravity and superstring corrections would only be a consistent approximation for 
$\alpha \ll 1$.  However, this limit is difficult to
analyze numerically. We will present numerical
results for $\alpha=1$ and argue separately on the effect of lowering
$\alpha$.  

To study straight vortices along, say, the $z-$direction,
we drop the z dependence and set $A_z=0$.  For time independent
configurations and defining $\tilde S=s+2i\alpha^2 a$, and $\tilde
D_\mu \tilde S = \partial_\mu \tilde S +2i\alpha^2 A_\mu$, the energy
functional can be written in the Bogomolnyi form:
\begin{eqnarray}
\mathscr{E} (\xi g)^{-2} & = &  |(D_x \pm iD_y)\phi |^2 +\scriptstyle 
\frac{1}{4} \displaystyle (\alpha s)^{-2} | ( \tilde D_x \pm i \tilde D_y) \tilde S|^2 {}
\nonumber \\
& &{} +\scriptstyle \frac{1}{2} \displaystyle 
(F_{x y} \mp (1-s^{-1}- |\phi|^2))^2 \pm  F_{x y}{}
\nonumber \\
&& {} \mp i[\partial_x(\phi^*D_y \phi)-\partial_y (\phi^*D_x \phi)]{}
\nonumber \\
&& {} \pm i \scriptstyle \frac{1}{2} \displaystyle \alpha^{-2}[\partial_x( s^{-1} 
\tilde D_y \tilde S) - \partial_y (s^{-1} \tilde D_x \tilde S)] .
\label{Bogonyi_nrgy2}
\end{eqnarray}
leading to a  bound on the energy per unit length
\begin{equation}
E = \scriptstyle \int \displaystyle d^2x \mathscr{E} \ge (\xi g)^2
\scriptstyle \int \displaystyle d^2x F_{x y}.
\end{equation}
The bound is attained by the solutions of the Bogomolnyi equations
\begin{eqnarray}
(D_x \pm iD_y)\phi &= 0 \nonumber \\
( \tilde D_x \pm i \tilde D_y) \tilde S &=0 \nonumber \\
F_{x y} \mp (1-s^{-1}- |\phi|^2) &=0 
\end{eqnarray}

We will focus on cylindrically symmetric vortices, that can be
described by the following ansatz:
\begin{eqnarray}
\phi = f(r) e^{i n \theta} & \qquad s^{-1} = h^2 \nonumber \\
a=m \theta  \qquad & \qquad A_\theta = v(r) / r  \ .
\end{eqnarray} 
(The ansatz for the dilaton makes comparison to the semilocal case
easier. Also we shall see that $h$ vanishes in various cases, and with
this choice we avoid having to deal with infinities).
 
With this ansatz the energy density becomes:
\begin{equation}
\mathscr{E} =A(r)^2 +  B(r)^2/(\alpha \, h)^{2}+\scriptstyle \frac{1}{2} 
\displaystyle C(r)^2+ D(r)'/r
\label{e_bog}
\end{equation}
and the Bogomolnyi equations
\begin{eqnarray}
A(r)&\equiv&f'-f(|n| - v)/r = 0\nonumber \\
B(r)&\equiv&h'-\alpha^2 h^3(|m| - v )/r = 0 \\ 
C(r)&\equiv&v'/r-( 1 - f^2 - h^2 )= 0  \nonumber \\
D(r) &\equiv& v -  f^2(v-|n|)-h^2(v-|m|)= 0
\label{bog}
\end{eqnarray}

The total energy is:
\begin{equation}
E=2 \pi \scriptstyle \int \displaystyle  d r r \mathscr{E}(r) \ge 2 \pi v_\infty 
\label{E_bog_nrgy}
\end{equation}
where $v_\infty$ is the asymptotic value of the gauge field profile function 
for large values of $r$. 

The condition $A(r) = B(r) = 0 $ implies the following relation
between the tachyon and the dilaton:
\begin{equation}
1/(\alpha \, h)^2 = 2(|n|-|m|)\log r - 2\log f + \kappa. 
\label{dilat_eq}
\end{equation}

Depending on the asymptotic value of the profile  functions for large 
values of $r$, $f_{\infty}$, $h_{\infty}$ and $v_{\infty}$, we can classify 
the solutions to the Bogomolnyi equations in three different families. Each 
of them is parametrized by the integration constant  $\kappa$.\\

$\bullet$ In the first case, 
\begin{equation}
f_{\infty} = 1, \quad h_{\infty}=0, \quad v_{\infty} = n \ , 
\end{equation}
the tachyon acquires a non vanishing vacuum
expectation value far from the center of the string.
The magnetic flux of these vortices is induced by the winding of the
tachyon, $n$.

The profile function $h(r)$ tends very slowly, (logarithmically), to
zero at large $r$. The details about the asymptotics of the fields
can be found in \cite{BDR}. Following Blanco-Pillado et al. we call
these vortices $\phi$-strings.\\
 
\begin{figure} 
\centering 
\includegraphics[width=0.5\textwidth]{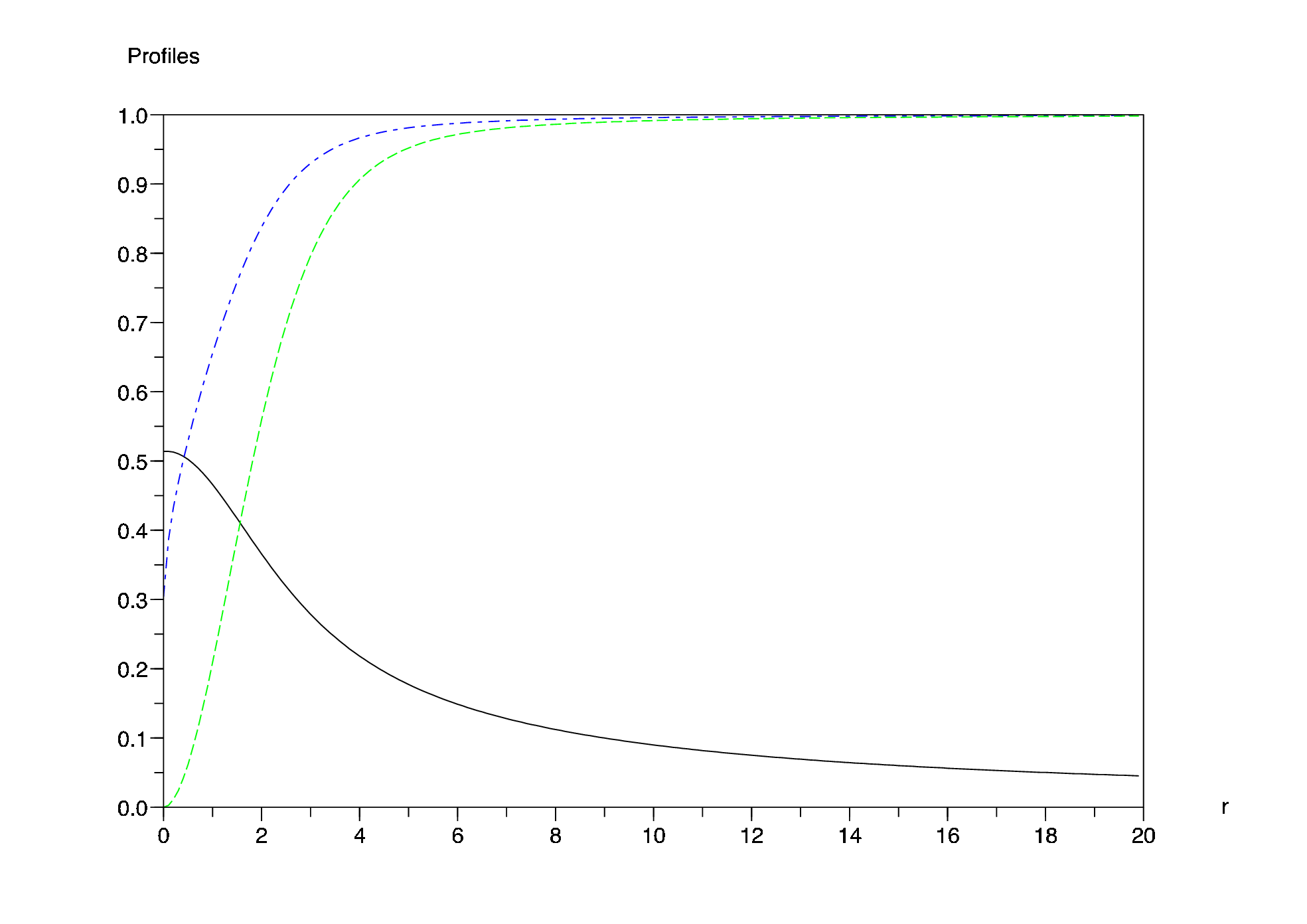}
\caption{The axion string profile functions, for m=1, n=0, $a(r)$, 
(upper dashed line), $h(r)$, (middle dashed line) and $f(r)$, (lower solid line), 
with the size of the condensate, $f_0=0.51$.}   
\label{fig2}
\end{figure} 

$\bullet$ In the second case, Fig.(\ref{fig2}),
the dilaton alone is responsible for compensating the D-term. The
function $h(r)$, approaches a non vanishing constant far from the
centre of the string, while the tachyon expectation value tends to
zero. In this family the magnetic flux is induced by the winding of
the axion, $m$.
\begin{equation}
f_{\infty} = 0 \quad h_{\infty}=1  \quad v_{\infty}= m  
\end{equation} 
These vortices have been denominated $s$-strings. They are regular
thanks to the constant Fayet-Iliopoulos term\\

In the previous two families the constant $\kappa$ has the same interpretation.
Each value of this parameter  is associated to a particular width of the strings.\\

$\bullet$ For strings of the third family both the tachyon
and the dilaton contribute to compensate the $D$-term.   This happens
when the axion and dilaton have the same winding.
\begin{equation}
f_{\infty}^2 + h_{\infty}^2 =1 \qquad v_{\infty} = n=m  
\label{mixed}
\end{equation}   
In this case $f_{\infty}$ and $h_{\infty}$ can have any value as long as the previous 
relation is satisfied (\ref{mixed}). Each particular $f_{\infty}$ can be associated to 
a single $\kappa$, which means that the interpretation of this parameter is different 
to the previous two cases, and cannot be related any more to the width of the strings, 
so we will not discuss it any further.\\ 

As can be seen in (\ref{bog}) the derivatives of $h$ scale as
$\alpha^2$.  In the case of the $\phi$-strings, varying $\alpha$ for a
fixed value of the integration constant k does not change much the width
of the string, but the condensate flattens. In the limit when
$\alpha$ is very small the $\phi-$strings are similar to a
Nielsen-Olesen string.  In the case of $s-$strings the width of the
string increases with decreasing $\alpha$, while the condensate
does not  vary much.  In fact, the main effect of decreasing $\alpha$
will be a slowing down of the dynamics.

\section{Discretized Equations of Motion}

The functions $f(r)$, $h(r)$ and $v(r)$ are substituted by the set of quantities
$f_k$, $h_k$ and $v_k$ which are the profile functions evaluated in the lattice 
points $r_k=(k+\scriptstyle \frac{1}{2} \displaystyle)\Delta$, where $\Delta$ 
is the lattice spacing. \\

We want to analyze the response of the solutions of the Bogomolnyi
equations under perturbations. We must make sure that the
configurations that we are going to perturb are stationary solutions
of the equations of motion. This is automatic in the continuous case
but not in an arbitrary discretization.\\

Following Leese \cite{leese} we construct a discrete version of the energy functional for static configurations given by:
{\setlength\arraycolsep{.5pt}
\begin{eqnarray}
\mathscr{E}_k  \Delta^2  & = & A_k^2 + \frac{4 B_k^2}{\alpha^2 (h_{k+1}+h_k)^2}+ \frac{C_k^2}{2 \Delta^2} 
+\frac{(D_{k+1}-D_k)}{k+ \scriptstyle \frac{1}{2}} \ 
\label{dBogEnergy}
\end{eqnarray}}
where
\begin{eqnarray}
A_k & = & f_{k+1} - f_k - (n-v_k) \frac{f_{k+1}+f_k}{2 k +1},  \nonumber \\
B_k & = & h_{k+1} - h_k - \alpha^2 (m-v_k) \frac{(h_{k+1}+h_k)^3}{2 k +1}, \nonumber \\
C_k & = &  \frac{ v_{k+1} -v_k}{k+ \scriptstyle \frac{1}{2}}-\Delta^2(1-h_{k+1}^2-f_{k+1}^2), \nonumber \\
D_k &=& v_k-f_k^2 (v_k-n)-h_k^2 (v_k-m).
\end{eqnarray}

The profile functions are obtained minimizing the total energy for static configurations:
\begin{equation}
E = 2 \pi \sum_{k=0}^{\infty} \Delta^2 (k+\scriptstyle \frac{1}{2}) \displaystyle \mathscr{E}_k, 
\label{BogEnergy}
\end{equation}
for which the discretized Bogomolnyi equations have to be satisfied: $A_k = B_k = C_k =0$.\\
 
The boundary conditions used to solve them are set at the points
$r_0$ and $r_1$.  At $r_0$ we impose $v_0=0$, and at $r_1$ we fix
the values of $f_1$ and $h_1$.  One of these two has to be tuned in
order to obtain the correct asymptotic behavior, and each value of
the other one corresponds to a different solution within a family.\\

The following discretized version of the action is naturally associated to the energy 
functional (\ref{BogEnergy}):
{\setlength\arraycolsep{1pt}  
\begin{eqnarray}
S &=& -2 \pi \tau \Delta^2 \sum_{l,k=0}^\infty (k+ \scriptstyle \frac{1}{2} \displaystyle )
 (\mathscr{E}_k^l-\mathscr{T}_k^l)
\nonumber \\
\tau^2 \mathscr{T}_k^l&=& (f_k^{l+1}-f_k^l)^2+ \frac{(h_k^{l+1}-h_k^l)^2}{\alpha^2 h_k^2}  -  
\frac{(a_k^{l+1}-a_k^l)^2}{(2 k+1) \Delta^4 } \ 
\label{action}
\end{eqnarray}}

The superscript $l$ labels the time slices, which are separated by an interval $\tau$.\\
Here $\mathscr{T}_k^l$ represents the density of energy associated to the time derivatives.  

The equations of motion can be derived from (\ref{action}) by setting
to zero the partial derivatives of the action with respect to the variables
$f_k^l$, $h_k^l$, $v_k^l$.\\

As the solutions of the Bogomolnyi equations are static and minimize the discretized
energy functional they must be stationary solutions of the discretized time dependent 
equations of motion. \\

The boundary conditions have been implemented using the same method of
\cite{leese}.  All the quantities measured take information from a
region of radius $r_{cal}=4.5$.  The simulation is stopped at
$t_{max}=2(r_{max}-r_{cal})$. In this way the region from which we
take data is not affected by the presence of the boundary.  A
typical value for the size of the lattice is $r_{max}=21.5$, but it
varies depending on the profile.\\

During the simulation we keep track of the following quantities:
\begin{eqnarray}
E(l,k_{cal})&=&2 \pi \sum_{k=0}^{k_{cal}} \Delta^2 (k+\scriptstyle \frac{1}{2}) \displaystyle 
\mathscr{E}_k^l\\ 
T(l,k_{cal})&=&2 \pi \Delta^2 \sum_{k=0}^{k_{cal}} (k+\scriptstyle \frac{1}{2} \displaystyle) 
\mathscr{T}_k^l\\
E_T(l,k_{cal})&=& \frac{E(l,k_{cal}) + T(l,k_{cal})}{E(0,k_{cal}) + T(0,k_{cal})}\\
W &=& 2 \pi \Delta^3  \frac{\sum_{k=0}^{k_{cal}}(k+\scriptstyle \frac{1}{2} \displaystyle)^2 \mathscr{E}_k^l}
{E(l,k_{cal})} \label{width}\\
F(l,k_{cal}) &=& 2 \pi (D_{k_{cal}}^l-D_0^l)
\label{flux}
\end{eqnarray}
Here $k_{cal}$ is defined by $r_{cal}=(\scriptstyle \frac{1}{2}
\displaystyle + k_{cal}) \Delta$.\\ 
$E$ is the static energy contained in $r<r_{cal}$, and $T$ the energy due to the time derivatives in the same region. $E_T$ is the total energy normalized to the initial value. $W$ is a measure of the width of the string. In (\ref{width}) the expression in the numerator is similar to the energy functional, but the extra factor $(k+\scriptstyle \frac{1}{2} \displaystyle)$ gives more weight to the energy far away from the core. Then, $W$ increases as the energy spreads. $F$ gives the magnetic flux confined in the region $r<r_{cal}$. \\ 

To implement the perturbation, we set a solution of the discrete
Bogomolnyi equations in the first time slice $l=0$, and in the second, $l=1$,
we put the same solution slightly deformed.
We characterize the strength of the perturbation by the fractional change of width between the first, ($l=0$), and second, ($l=1$), time slices:
\begin{equation}
\dot W_0 \equiv (W(l=1) - W(l=0))/W(l=0).
\end{equation} 
During the simulation we have set $\Delta=0.1$ and $\tau=0.05$, with 
$\tau/\Delta$ smaller than the Courant number $1/\sqrt{2}$.

\begin{figure} 
\centering \includegraphics[width=0.5\textwidth]{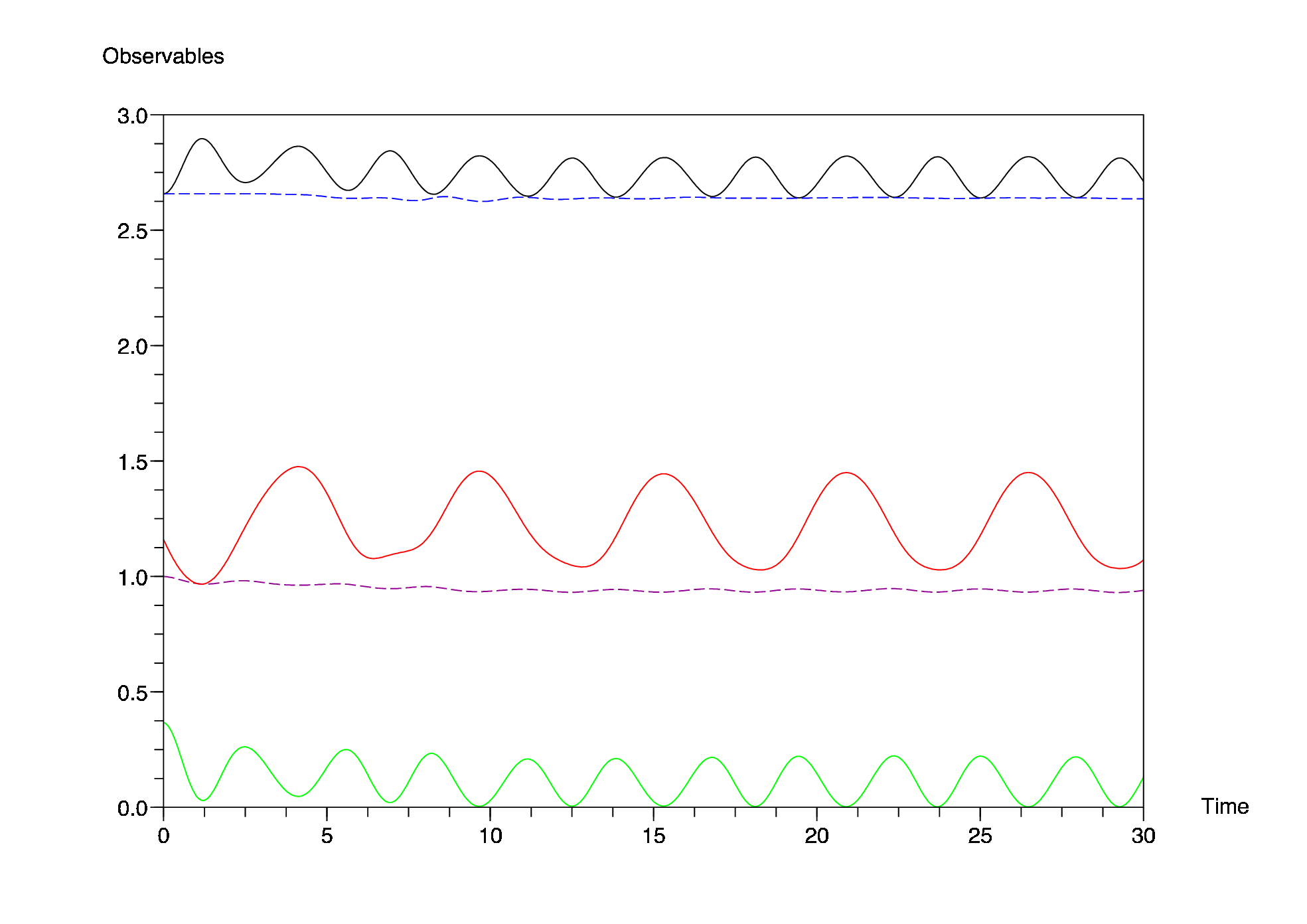}
\caption{Response of a $\phi-string$ ($n=1,\ m=0$), with $\alpha=1$
and condensate size $h_1=0.5$, to a perturbation with strength $\dot
W_0=-0.261$. The plotted lines are, from top to bottom, $E$, $F$, $W$,
$E_T$ and $T$ (except for $W$ and $E_T$, the rest of the plots have
been rescaled by a factor of $1/2$ to fit in the window). The core
width, $W$, oscillates but is constant on average showing that the
zero mode is not excited.}
\label{fig4}
\end{figure}

\subsection{Tachyonic Strings}
In this case to implement the perturbation we take the second time slice to be :
\begin{eqnarray}
f_k^1=(1 + p(r)) f_k^0, \quad a_k^1= (1 + p(r)) a_k^0
\nonumber \\
h_k^1=( \; 1/(h_k^0)^2+2\log( \, 1 - p(r) \, ) \; )^{-1/2}
\label{pert1}
\end{eqnarray}
with the perturbation $p(r)=p_0[1-3(r/r_0)^2+2(r/r_0)^3]$ for $r< r_0$
and zero otherwise.\\
The perturbation has been chosen in order to maximize 
the fraction of energy absorbed by the zero mode. Notice the relation between 
the perturbations of the tachyon field and the dilaton,
and the equation (\ref{dilat_eq}) that gives $h$ in terms of $f$, in fact for small values of the parameter $p_0$ the perturbed profile approximately satisfies the Bogomolnyi equations. 
This perturbation gives a coordinate dependence to the integration constant $\kappa$.\\
 The results shown are for a value of $p_0$ such that the 
perturbation initially reduced the width of the string, but the same results are 
obtained in the opposite case.
The parameter $r_0$ varies for different initial conditions. 
In general it has a value close to $r_0=2$.\\

Fig.(\ref{fig4}) shows the evolution of a $\phi$-string, with $n=1$,
$m=0$ and a core size $h_1=0.51$.  We show the case $\alpha=1$,
which is also the choice made in \cite{BDR}.  The perturbation applied
has a strength $\dot W_0= -0.261$, which corresponds to a $0.4 \%$
perturbation in the energy.  We have plotted the observables defined
in the last section as a function of time.  Upper solid line
represents $E$, the dashed line just below is the magnetic flux in
$r<r_{cal}$, which remains almost constant. The kinetic energy, $T$,
is represented by the bottom solid line. 
The dashed line at the center of the figure is the total energy.  
Although it can not be clearly seen in the 
plot, the data tell us that during the period before
$t=12$, a fraction of the energy is lost.  This is the
initial burst of radiation emitted after the perturbation.  After that
the system reaches a stationary state where all quantities oscillate except the
magnetic flux and the total energy.\\

The remaining solid line in the center is the string width.  
As the energy contained in the region $r\le r_{cal}$ remains 
constant and the width of the vortex oscillates only around the 
initial value we conclude that this kind of string is stable under 
this perturbation.  The experiment has been repeated in a wide range 
of the parameters, $p_0$, $r_0$, and for different initial widths 
(parametrized by $h_1$), and windings, but the results are similar 
to the ones presented here.  Nielsen-Olesen vortices react in the 
same way to a perturbation, as was shown in \cite{leese}.\\

We have repeated the evolution for different values of $\alpha$ but no
qualitative change has been observed. As was mentioned before, the
smaller the value of $\alpha$, the more similar the $\phi-$string is
to a regular Nielsen-Olesen string, which is known to be stable.

\begin{figure} 
\centering \includegraphics[width=0.54\textwidth]{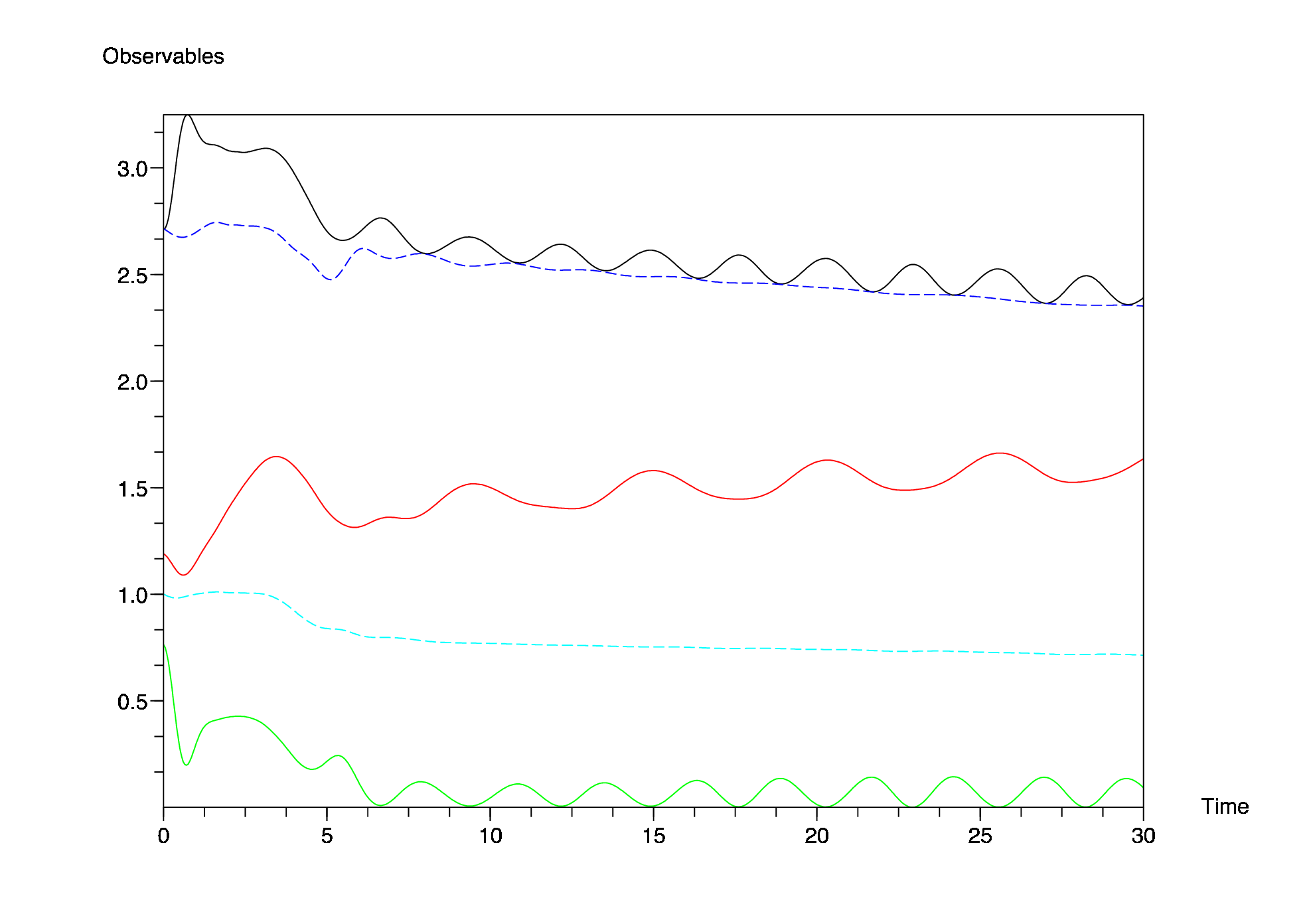}
\caption{Response of an $s-$string ($n=0$, $m=1$), with $\alpha=1$ and
condensate size $f_1=0.49$, to a perturbation with strength $\dot
W_0=-0.095$. The plotted lines are, from top to bottom, $E$, $F$, $W$,
$E_T$ and $T$ (except for $W$ and $E_T$, the rest of the plots have
been rescaled by a factor of $1/2$ to fit in the window). The core
width, $W$, after a transient, oscillates and increases at a constant
rate, the zero mode is excited in this case.}
\label{fig5}
\end{figure} 

\subsection{Axionic Strings}
The relevant perturbation that excites the zero mode  in this case is:
\begin{eqnarray}
f_k^1=(1- p(r)) f_k^0, \quad a_k^1= (1 -p(r)) a_k^0
\nonumber \\
h_k^1=( \; 1/(h_k^0)^2+2\log( \, 1 + p(r) \, ) \; )^{-1/2}
\label{pert2}
\end{eqnarray}

The result of applying this perturbation with a strength of $\dot
W_0=-0.095$ to an $s$-string, with $\alpha=1$, can be seen in Fig.({\ref{fig5}). In this case the perturbation in energy is $2.4 \%$. 
The string has windings $n=0$ and $m=1$, and core size $f_1=0.51$.
The functions plotted are the same ones that appear in Fig.(\ref{fig4}).  One
of the most relevant features of these plots is that the magnetic
flux and the total energy are decreasing with time, which implies that
the energy is flowing out from the region $r \le r_{cal}$, and the
magnetic flux is spreading. At the same time the width of the string,
ignoring the oscillatory behavior, increases at a constant rate.
Before $t=5$ oscillations are noisy. In this period the shock wave
produced by the perturbation is still inside $r \le r_{cal}$.\\

Although the perturbation was chosen to reduce the core width, the
time interval when the core is contracting cannot be seen clearly in
the figures.  The reason is that the contracting regime ends before
the initial burst of radiation comes out from the observed region. As
the system is not in a steady state yet, the data are difficult to
interpret.  We have chosen to show this case because the expanding
regime is shown more clearly.\\

\begin{figure}[h] 
\centering \includegraphics[width=0.5\textwidth]{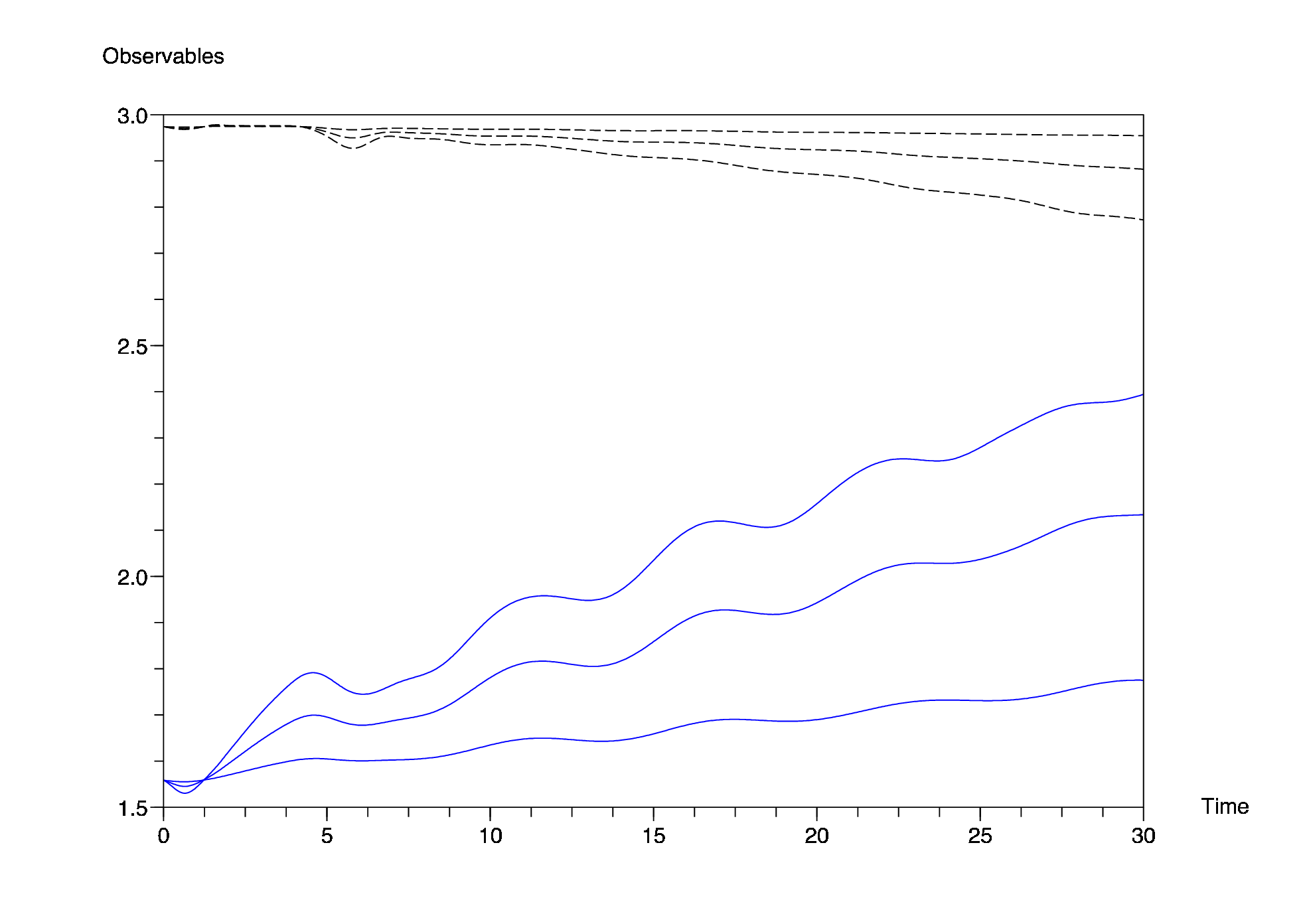}
\caption{Response of an $n=0$, $m=2$, $\alpha=1$ $s$-string with a
core condensate of size $f_1=0.51$ to perturbations with different
strengths. Dashed lines correspond to $F$, the strength is lowest for
the top one. Solid lines represent $W$, the strength is highest
for the top one. The strengths are: $\dot W_0 = -0.004$, $ -0.014$ and
$-0.024$. $F$ is rescaled by a factor of $1/4$. The figure shows how
the growth rate of the core increases with the perturbation
strength. }
\label{fig6}
\end{figure}

Fig.(\ref{fig6}) shows the effect of applying perturbations of
different strengths to a $n=0$, $m=2$ vortex. In this case the vortex
has also a condensate size of $f_1=0.51$. The strengths applied are:
$\dot W_0 =-0.004$, $\dot W_0 =-0.014$ and $\dot W_0 =-0.024$.\\
  
Notice that the bigger the strength of the perturbation, the larger
the fraction of magnetic flux lost through the boundary $r=r_{cal}$. The
rate of growth of the radius also increases with the strength of the
perturbation.\\

In Fig.(\ref{fig7}) it can be seen how the rate of expansion of an
$s-$string, ($m=1$, $n=0$), is affected by varying the value of
$\alpha$. In this case the perturbation has been chosen to initially
increase the core size $\dot W_0 = 0.233>0$.  As we decrease $\alpha$,
keeping the perturbation strength fixed, the rate of expansion of the
string decreases.  This can be understood from equation
(\ref{action}). The energy associated to the field $h$ scales as the
inverse of $\alpha^2$. Deviations from the solution to the Bogomolnyi
equations cost more energy for smaller values of $\alpha$, thus for a
fixed perturbation strength the evolution rates should decrease with
$\alpha$.  The values of alpha are: $\alpha=0.95$, $0.94$, $0.90$. \\

The precision of the technique used here does
not allow to obtain reliable data for values of $\alpha$ lower than
$0.7$, where already the evolution is so slow that it can hardly be
appreciated during the time of the simulation. However, note that in
these simulations time is measured in units of the inverse of the
Higgs mass, thus even for lower values of $\alpha$ 
decompactification is still possible  on cosmological time scales. \\

\begin{figure}[t]
\centering \includegraphics[width=0.5\textwidth]{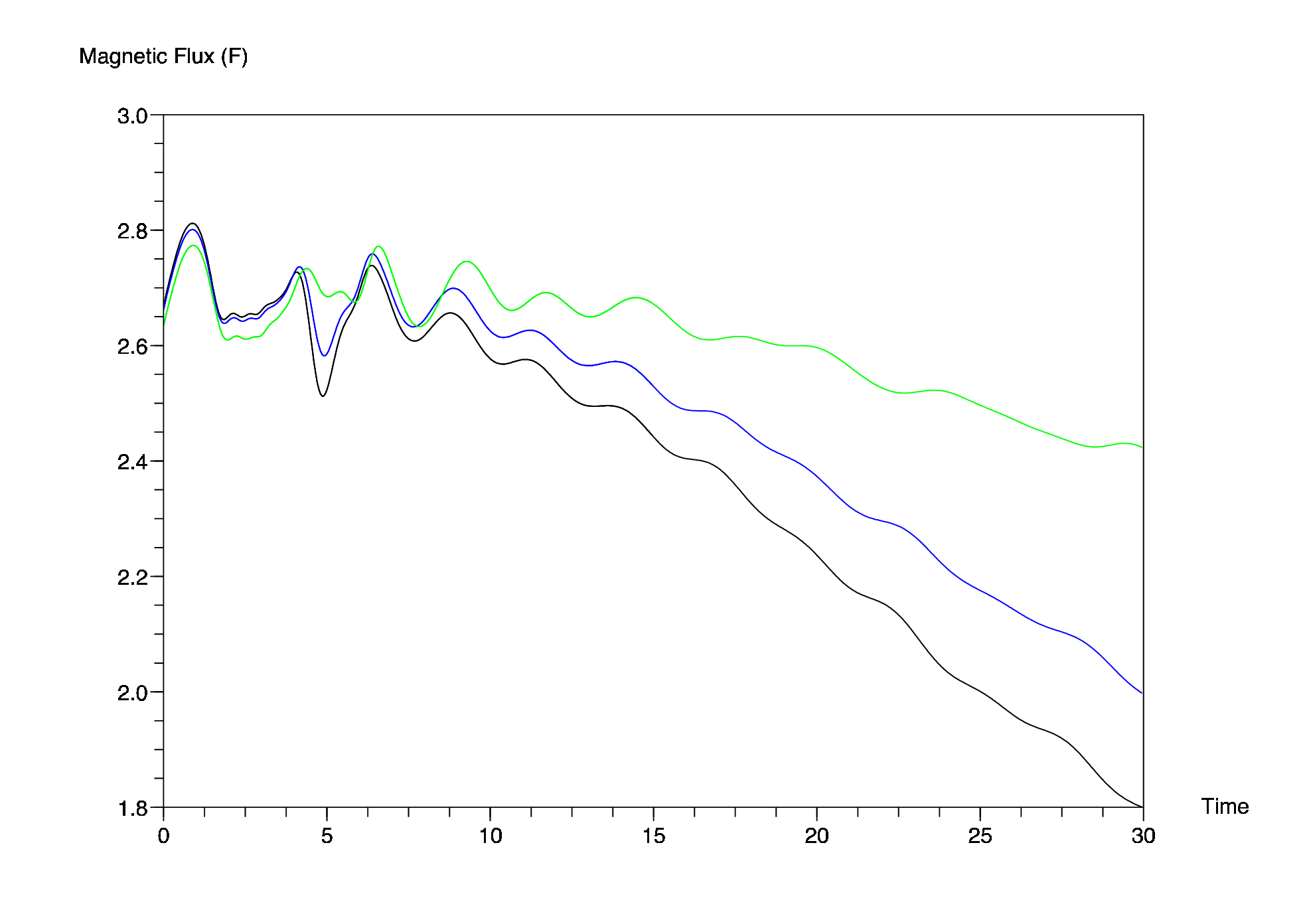}
\caption{Response of an $n=0$, $m=1$ $s$-string with a core condensate
of size $f_1=0.51$ to a perturbation with $\dot W_0 =0.233$. The
curves represent $F$, the magnetic flux rescaled by a factor of $1/4$.
From bottom to top the values of $\alpha$ are are: $\alpha = 0.95$,
$0.94$ and $0.90$. The figure illustrates the slowing down of the core
expansion with decreasing $\alpha$}
\label{fig7}
\end{figure}

We are grateful to Jose Blanco-Pillado, Stephen Davis, Koen Kuijken
and Jon Urrestilla for very useful discussions. 
This work is supported
by Basque Government grant BF104.203, by the Spanish Ministry of
Education under project FPA 2005-04823 and by the Netherlands
Organization for Scientific Research (N.W.O.) under the VICI
programme. We are also grateful to the ESF COSLAB Programme for
incidental support in the initial stages of this work.\\

\end{document}